\documentclass[conference]{IEEEtran}
\IEEEoverridecommandlockouts
\usepackage{cite}
\usepackage{amsmath,amssymb,amsfonts}
\usepackage{algorithmic}
\usepackage{graphicx}
\usepackage{textcomp}
\usepackage{xcolor}
\usepackage{url}
\usepackage{subfig}
\usepackage{tikz}
\usetikzlibrary{patterns}
\usepackage{gensymb}
\usepackage{listings}
\usepackage{xcolor}
\usepackage{array}
\usepackage{bm}
\usepackage{colortbl}
\usepackage{wrapfig}

\newcommand{\system}{Northlight}

\newcommand{\smalltt}[1]{{\texttt{\small #1}}}
\newcommand{\scriptsizett}[1]{{\texttt{\scriptsize #1}}}

\def\BibTeX{{\rm B\kern-.05em{\sc i\kern-.025em b}\kern-.08em
    T\kern-.1667em\lower.7ex\hbox{E}\kern-.125em}}

\lstdefinestyle{sql}{
  frame=tb,
  language=SQL,
  morekeywords={extension},
  deletendkeywords={TIME},
  aboveskip=2mm,
  belowskip=2mm,
  captionpos=b,
  showstringspaces=false,
  columns=flexible,
  basicstyle={\footnotesize\ttfamily},
  numbers=none,
  numberstyle=\small\color{gray},
  keywordstyle=\color{blue},
  commentstyle=\color{magenta},
  frame=none,
  breaklines=true,
  breakatwhitespace=true,
  tabsize=3,
}

\lstdefinestyle{pseudo}{
  frame=tb,
  language={scala},
  deletekeywords={with},
  aboveskip=2mm,
  belowskip=2mm,
  captionpos=b,
  showstringspaces=false,
  columns=flexible,
  basicstyle={\scriptsize\ttfamily},
  numbers=left,
  numberstyle=\tiny \color{black},
  keywordstyle=\color{blue},
  commentstyle=\color{magenta},
  frame=none,
  breaklines=true,
  breakatwhitespace=true,
  tabsize=3,
}

\renewcommand{\emph}[1]{\textit{#1}}

\definecolor{jgu}{RGB}{193,0,42}
\definecolor{grayshade}{RGB}{180,180,180}
\definecolor{lightgrayshade}{RGB}{230,230,230}

\makeatletter
\newcommand{\linebreakand}{%
  \end{@IEEEauthorhalign}
  \hfill\mbox{}\par
  \mbox{}\hfill\begin{@IEEEauthorhalign}
}
\makeatother

\title{\system{}: Declarative and Optimized Analysis of Atmospheric Datasets in SparkSQL}

\author{\IEEEauthorblockN{Justus Henneberg}
\IEEEauthorblockA{\textit{Institute of Computer Science} \\
\textit{Johannes Gutenberg-University}\\
Mainz, Germany\\
henneberg@uni-mainz.de}
\and
\IEEEauthorblockN{Felix Schuhknecht}
\IEEEauthorblockA{\textit{Institute of Computer Science} \\
\textit{Johannes Gutenberg-University}\\
Mainz, Germany \\
schuhknecht@uni-mainz.de}
\and 
\IEEEauthorblockN{Philipp Reutter}
\IEEEauthorblockA{\textit{Institute of Atmospheric Physics} \\
\textit{Johannes Gutenberg-University}\\
Mainz, Germany \\
preutter@uni-mainz.de}
\linebreakand
\IEEEauthorblockN{Nils Brast}
\IEEEauthorblockA{\textit{Institute of Atmospheric Physics} \\
\textit{Johannes Gutenberg-University}\\
Mainz, Germany \\
nibrast@uni-mainz.de}
\and
\IEEEauthorblockN{Peter Spichtinger}
\IEEEauthorblockA{\textit{Institute of Atmospheric Physics} \\
\textit{Johannes Gutenberg-University}\\
Mainz, Germany \\
spichtin@uni-mainz.de}
}

\begin{document}

\maketitle

\begin{abstract}

Performing data-intensive analytics is an essential part of modern Earth science. As such, research in atmospheric physics and meteorology frequently requires the processing of very large observational and/or modeled datasets. Typically, these datasets (a)~have high dimensionality, i.e. contain various measurements per spatiotemporal point, (b)~are extremely large, containing observations over a long time period. Additionally, (c)~the analytical tasks being performed on these datasets are structurally complex. 

Over the years, the binary format NetCDF has been established as a de-facto standard in distributing and exchanging such multi-dimensional datasets in the Earth science community -- along with tools and APIs to visualize, process, and generate them. Unfortunately, these access methods typically lack either (1)~an easy-to-use but rich query interface or (2)~an automatic optimization pipeline tailored towards the specialities of these datasets. 
As such, researchers from the field of Earth sciences (which are typically not computer scientists) unnecessarily struggle in efficiently working with these datasets on a daily basis. 

Consequently, in this work, we aim at resolving the aforementioned issues. Instead of proposing yet another specialized tool and interface to work with atmospheric datasets, we integrate sophisticated NetCDF processing capabilities into the established SparkSQL dataflow engine -- resulting in our system~\system{}. In contrast to comparable systems, \system{} introduces a set of fully automatic optimizations specifically tailored towards NetCDF processing. We experimentally show that \system{} scales gracefully with the selectivity of the analysis tasks and outperforms the comparable state-of-the-art pipeline by up to a factor of~6x.
\end{abstract}

\begin{IEEEkeywords}
NetCDF, SparkSQL, Query Optimization, Earth Science, Atmospheric Physics
\end{IEEEkeywords}

\section{Introduction}
\label{sec:intro}

Modern research in the field of Earth science has to process large volumes of observational/modeled data. Also, the joint processing and comparison of data from different sources is a common task in Earth science research~\cite{lit:rhi, Gierens:1999tf}, further increasing the total amount of data to process. For instance, in the field of atmospheric physics and meteorology, researchers perform an analysis on the ERA5~\cite{lit:ecmwf-era5,lit:era5} dataset provided by the European Centre for Medium-Range Weather Forecasts, which contains around 5~PB of multi-dimensional climate data estimates since~$1979$. Processing such a large and potentially distributed dataset efficiently is a non-trivial task.
Further, the datasets are typically split across a large number of files, where certain dimensions might span over multiple files, while others remain consistent for each file.  
Handling the aforementioned challenges manually is especially undesirable for users who originate from a domain other than computer science. Researchers from natural sciences, who want to focus on their actual task, are forced to think about efficient data management and how to carefully optimize complex query plans in order to achieve acceptable runtimes. 

\subsection{\system{}}

As a consequence, in the following work we propose the system~\system{}, which combines an easy-to-use declarative query interface with high processing performance, specifically tailored for atmospheric datasets. 

For decades, the distribution of observational/modeled data has been a cornerstone in atmospheric physics, meteorology and weather prediction, using the \textit{NetCDF}~\cite{lit:netcdf} format as a de-facto standard for more than $30$~years. Thus, we focus on datasets materialized in the NetCDF format in the following.
At the core, it is a self-describing, multi-dimensional binary format with a focus on portability and scalability. Interestingly, while there are plenty of low-level APIs available for creating, modifying, and accessing NetCDF files, there is hardly any NetCDF-capable processing system with a rich query interface that combines convenience with performance. This is especially surprising considering the typical userbase of the format.

Instead of proposing yet another specialized tool for NetCDF processing that requires the users to adopt to, we integrate NetCDF support into the well known \textit{SparkSQL}\cite{lit:spark-sql,lit:spark-sql-diss} framework. SparkSQL provides exactly what researchers from the Earth science community require: On the one hand, they need an abstraction layer which allows to easily connect NetCDF datasets and formulate analysis tasks in a simple declarative language such as SQL. On the other hand, they can be sure that their query is automatically and transparently optimized into an efficient parallel and distributed execution plan. 

In summary, our work makes the following contributions:

\subsection{Contributions}

\textbf{(1)~Atmospheric Datasets in SparkSQL}: We integrate rich support for processing distributed NetCDF datasets into SparkSQL. Users are able to formulate queries against a convenient row-wise representation of their NetCDF dataset using SQL. \system{} ensures that the formulated query is automatically and transparently optimized down to the data source. \system{} seamlessly extends an existing Spark\cite{lit:spark,lit:spark-diss} installation -- no modifications to the vanilla Spark code or the optimization pipeline is required. This allows an easy installation and adoption of our techniques for users of both Spark and SparkSQL. The full source code of the system will be available upon acceptance. 

\noindent \textbf{(2)~Transparent Query Optimization}: We propose several fully automatic data management optimizations to let the SparkSQL pipeline efficiently handle the characteristics of NetCDF datasets. These optimizations include vertical and horizontal pruning, which happens directly at the data source. We do so without precomputing any auxiliary index structure. 

\noindent \textbf{(3)~Optimizing Non-Convex Predicates}: Besides supporting the optimization of simple convex predicates, like done in comparable systems, we also target the optimization of complex non-convex predicates. These consist, for instance, of several conjuncted areas that might contain holes or overlap with each other. Our optimizations ensure that such non-convex predicates, which occur frequently in atmospheric analysis tasks, still result in loading the minimum amount of required data from the data source.

\noindent \textbf{(4)~Optimizing Joins via Envelopes}: Further, we support a transparent optimization of certain join queries via the introduction of envelopes. Envelopes are automatically computed lower and upper bounds on the coordinate axes. These are injected into SparkSQL's optimizer without changing the optimization pipeline in any way and can drastically speed up joins on coordinates. 

\noindent \textbf{(5)~Evaluation of Real-world Applications}: We showcase how a set of real-world computations from the domain of atmospheric physics can be expressed and evaluated on real-world observational/modeled datasets within our framework. We show that \system{} gracefully scales with the selectivity of the workload for various workloads and optimizes join processing via envelopes. We compare our optimization pipeline to the one of ClimateSpark~\cite{lit:climatespark} and show that we optimize deeper without requiring any auxiliary data structures.




This paper is structured as follows: In Section~\ref{sec:background}, we discuss the required background, such as the structure of the NetCDF format, as well as the related work. The related work also motivates and steers our overall architectural design, which we will describe in Section~\ref{sec:architecture}. Based on that, in Section~\ref{ssec:convex_query_processing}, we discuss how query processing and optimization work when facing convex predicates, whereas in Section~\ref{ssec:non_convex_query_processing}, we discuss the significantly more complex non-convex case. In Section~\ref{sec:envelopes}, we target the optimization of specific joins using envelopes. In Section~\ref{sec:experiments}, we conclude with an extensive experimental evaluation. 

\section{Background}
\label{sec:background}

To motivate the approach we are taking, in the following we first briefly introduce the NetCDF file format, then present state-of-the-art tools/frameworks/systems and also discuss their limitations.

\subsection{The NetCDF File Format}
\label{sec:system:problem:netcdf}

\begin{figure}[h!]
\includegraphics[width=\columnwidth]{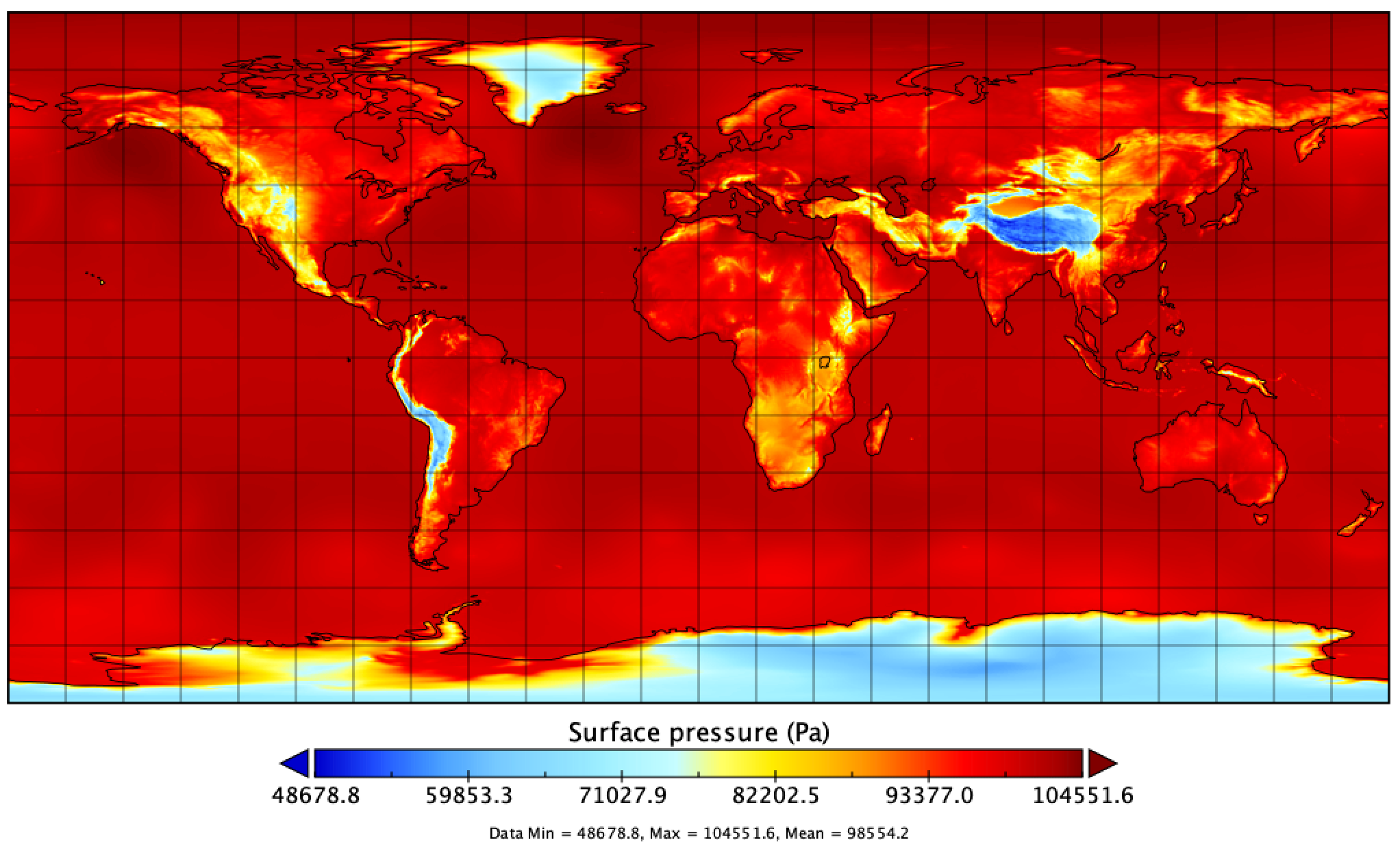}
\caption{Visualization of the \smalltt{sp} variable (surface pressure) of the ERA5 surface data subset in Panoply~\cite{lit:panoply}.}
\label{fig:netcdf_viz}
\end{figure}

As mentioned before, we target the efficient processing of \textit{NetCDF}~\cite{lit:netcdf} datasets. This self-describing columnar file format for multidimensional data has emerged as the de-facto standard format used by the Earth science research community to materialize and distribute atmospheric datasets, along with an extensive list of metadata conventions\cite{lit:convention,lit:convention-monotonic}.
A NetCDF file consists of multiple \textit{dimensions} and \textit{variables}. A variable stores an observation for each point on a fixed multi-dimensional grid. For example, our local surface-level subset of the ERA5 dataset consists of the three dimensions time, longitude, and latitude, and each of the $19$~variables maps instances of the tuple (\smalltt{time}, \smalltt{lon}, \smalltt{lat}) to the observation at these coordinates. Figure~\ref{fig:netcdf_viz} visualizes the variable \smalltt{sp} (surface pressure) at a single point in time along varying longitude and latitude. Additionally, each dimension can be accompanied by a so-called \textit{coordinate variable}. This coordinate variable describes the coordinate values of the dimension, e.g. that the \smalltt{lat} dimension runs from $-90\degree$ to $90\degree$ in quarter-degree steps.
Apart from that, NetCDF datasets are typically materialized as a set of individual files that must be processed conjunctively. 

Processing these datasets manually using the available low-level APIs~\cite{lit:netcdf-java,lit:netcdf-python,lit:xarray} imposes various difficulties on the user, in particular, if the user originates from a domain other than computer science. As a consequence, the community asks for frameworks that allow formulating complex queries in a declarative manner while ensuring the best possible performance during evaluation -- \system{} is our answer.

\subsection{Related Work}

While most NetCDF tools focus on small-scale data visualization~\cite{lit:panoply,lit:met3d,lit:cistools,lit:ncview} or conversion~\cite{lit:cdo}, a few existing tools support declarative queries against NetCDF datasets. Consequently, we discuss these in the following, as they also motivate our design in various stages: 

First of all, \emph{SciSpark}\cite{lit:scispark} extends the Spark framework with an RDD\cite{lit:rdd} implementation for multidimensional datasets. However, while SciSpark enables the distributed and parallel processing of NetCDF files, it lacks two crucial requirements: First, albeit mentioning Shark\cite{lit:shark} integration, there is no indication that SciSpark provides a high-level query interface. Therefore, the user is required to formulate the query against Spark's RDD interface, which requires deeper programming knowledge. Second, it does not apply any automatic optimizations, such as query rewriting, to the processing pipeline. This results in poor performance~\cite{lit:climatespark}, especially for selective queries. 

\textit{ClimateSpark}~\cite{lit:climatespark} goes one step further and integrates the support for processing multidimensional files into SparkSQL. Consequently, it utilizes the provided SQL interface and, at first glance, allows the convenient formulation of declarative queries that can be executed efficiently. However, a closer look reveals that various limitations are still imposed by the system, which drastically limit its usability: 
First, the provided SQL interface is restricted in several ways: Queries must factor in the reconstruction of actual tuples from the dimensions and variables of the multidimensional dataset. Furthermore, ClimateSpark only supports simple predicates that select a single convex region from the dataset. Frequently required compound predicates such as \smalltt{((lon > 0.0 AND lon < 20.0) OR (lon > 60.0 AND lon < 80.0))} either result in data being loaded unnecessarily or require a decomposition of the query at RDD level. In addition, overlapping conjunctive predicates cannot be handled at all by the engine.
Second, to speed up processing, ClimateSpark relies on a spatiotemporal index, which is materialized in a separate database. This index must be created for each variable of interest in advance to allow the pruning of data during loading. 
Third, ClimateSpark imposes certain requirements on the dataset: For example, only datasets that materialize one point in time per file are supported -- other datasets first require a manual conversion. 

\section{\system{}: Architectural Overview}
\label{sec:architecture}

In summary, the state-of-the-art systems do not satisfy the requirements of the community: We want to be able to (a)~formulate SQL queries against a convenient row-wise representation of the multidimensional dataset. (b)~We want to be able to use arbitrarily complex and potentially non-convex predicates. Still, the system should ensure to load the minimal amount of data from the data source.
(c)~We want to enable querying and optimization without any preprocessing of data. 


\subsection{Communicating with the NetCDF API}
\label{ssec:netcdf_api}

We utilize the official UCAR/Unidata NetCDF-Java library~\cite{lit:netcdf-java} to access NetCDF files. This library is based on Unidata's \textit{Common Data Model}\cite{lit:cdm} and can therefore handle both NetCDF-3 and NetCDF-4 files. To enable support for files stored in HDFS\cite{lit:hadoop-hdfs}, we open the files through a wrapper class\cite{lit:scispark-github} from SciSpark.
The NetCDF-Java library allows us to perform so-called \textit{subarray lookups} in a $d$-dimensional NetCDF file: By providing a multidimensional starting point along the dimensions, which is called the \textit{origin}~$o$, and a multidimensional length, called the \textit{extent}~$e$, the API returns a (multidimensional) subarray for a variable of interest.

It is important to note that a subarray lookup \textit{cannot} be done by passing concrete dimension values to the API, such as $(\smalltt{time}=01.01.2020, \smalltt{lon}=45.0\degree, \smalltt{lat}=12.25\degree)$. Instead, the API must receive a \textit{positional tuple} as the origin. For example, the positional tuple~$(\smalltt{time}=0, \smalltt{lon}=180, \smalltt{lat}=49)$ would refer to position~$0$ along the \smalltt{time} dimension, position~$180$ along the \smalltt{lon} dimension, and position $49$ along the \smalltt{lat} dimension. Accordingly, the extent describes the positional length for each dimension.
Together, origin and extent form a $d$-dimensional \textit{block}.

\subsection{Connecting NetCDF Datasets to SparkSQL}
With sufficient knowledge about the underlying API, let us now see how to connect a new NetCDF dataset to \system{}.
Essentially, the user only has to provide two things: 

\noindent \textbf{1)}~A set of NetCDF filenames, potentially distributed, representing the dataset to connect.

\noindent \textbf{2)}~For each dimension of the dataset, the information whether the dimension spans over multiple files or not. This information will become relevant to efficiently query the dataset. For example, in the ERA5 dataset, the \smalltt{time} dimension spans over multiple files: Each file contains exactly one point in time, representing a single hour. In contrast to that, the \smalltt{lon} and \smalltt{lat} dimensions do not span over multiple files: Each file contains the full $180\degree$~latitude and $360\degree$~longitude information. 

The schema of the dataset is automatically inferred by \system{}: We extract schema information about all dimensions and variables from the first file of the dataset, assuming the schema is shared by all files. The user can also specify a schema manually. 
Apart from schema inference, we also take into account how files are distributed across nodes. As a result, \system{} will allocate one RDD for the entire dataset, which contains one partition for each file. During execution, the scheduler of Spark will utilize this information when distributing tasks among the nodes of the cluster to execute them as locally as possible.

\subsection{NetCDF $\rightarrow$ RDD $\rightarrow$ Relation}
\label{ssec:netcdf_integration}

To represent our dataset internally, we utilize Spark's \textit{RDD}\cite{lit:rdd} as well as the \textit{Relations} concept from SparkSQL.
We use the RDD to load the dataset from disk and to convert it into a row-wise representation. As we expose one tuple per data point, users can conveniently operate on the dataset through the \textit{Dataset API}\cite{lit:datasets}.

The created row-wise representation follows the schema visualized in Table~\ref{table:schema}, using the ERA5 surface subset as an example: First, we expose the name of the file from which the data of the tuple originates. Then, for each dimension spanning multiple files, we expose a column with the dimension values (\smalltt{time}). Note that \system{} automatically converts the internal representation of \smalltt{time} into a user-friendly representation. For each non-spanning dimension, we expose a column with the dimension values and a column with the corresponding positions (\smalltt{lon} and \smalltt{lat}). The positional information is provided to allow advanced users to directly formulate queries against it. Finally, we expose a column for each variable.

\begin{table}[h!]
\setlength\tabcolsep{2.7pt}
\begin{tabular}{| c | c | c | c | c | c | c | c | c |}
\hline
\cellcolor{gray!25}\textbf{file} & \cellcolor{red!25}\textbf{time} & \cellcolor{green!25}\textbf{lon} & \cellcolor{green!25}\textbf{lonPos} & \cellcolor{green!25}\textbf{lat} & \cellcolor{green!25}\textbf{latPos} & \cellcolor{blue!25}\textbf{asn} & \cellcolor{blue!25}\textbf{blh} & \cellcolor{blue!25}...\\\hline\hline
04.nc & 2017-01-01 04:00:00 & 0 & 720 & 90.0 & 0 & 0.9 & 264.5 & ...\\\hline 
04.nc & 2017-01-01 04:00:00 & 0 & 720 & 89.75 & 1 & 0.9 & 333.8 & ...\\ \hline
04.nc & 2017-01-01 04:00:00 & 0 & 720 & ... & ... & ... & ... & ...\\  
\hline    
\end{tabular}
\caption{Exposed row-wise representation of ERA5 surface.}
\label{table:schema}
\end{table}

The Relation is responsible for exposing the schema of the data source, i.e.\ the data types of the columns, to the SparkSQL layer. The schema stores the name of each column, its corresponding data type, and whether the column is nullable. To obtain it, we convert the data types inferred earlier to the corresponding SparkSQL types.

With an overview of how we connect and represent distributed NetCDF datasets in SparkSQL, we can now discuss how our query processing layer works. In particular, we will discuss which optimizations we apply in order to prune data as close as possible to the data source when dealing with convex predicates (Section~\ref{ssec:convex_query_processing}) and non-convex predicates (Section~\ref{ssec:non_convex_query_processing}). Then, we discuss how to perform additional pruning in the presence of joins (Section~\ref{sec:envelopes}).

\section{Convex Query Processing \& Optimization}
\label{ssec:convex_query_processing}

We start the discussion by following the processing steps of a simple example query~$Q_1$, shown in Listing~\ref{listing:q1}. $Q_1$ loads a convex region of the previously mentioned ERA5 surface subset. Convex means that the selected region can be translated into a \textit{single} subarray lookup per file, represented by a single $d$-dimensional block.

\noindent 
\begin{minipage}{\linewidth}
\begin{lstlisting}[style=sql, caption={Example query~$Q_1$ (convex predicate).}, label={listing:q1}]
SELECT time, lat, lon, sp
FROM era_b
WHERE time > '2017-01-01 00:15:00' AND 
        lat > 20.2 AND lat < 60.5 
\end{lstlisting}
\end{minipage}

\noindent The query operates on the ERA5 surface subset, which is split into a set of files where each file stores all observations across longitude and latitude at a single point in time. More precisely, each file contains the following dimensions: \smalltt{time}~($1$~timestamp, resolution: hourly) $\times$ \smalltt{lat}~($721$~entries, resolution: $1/4\degree$) $\times$ \smalltt{lon}~($1440$~entries, resolution: $1/4\degree$). 
It is also noteworthy that the dataset may contain gaps, i.e. an hour of a day (and hence the corresponding file) might be simply missing.

\subsection{Overview}

Efficiently processing very large datasets relies on pruning all data that is not relevant for the query early on. Thus, fully scanning all files of the dataset, as done by SciSpark~\cite{lit:scispark} for instance, is not an option. In \system{}, we support both vertical and horizontal pruning, which is automatically pushed down from the SQL representation to the data source.

To implement vertical pruning, we filter and reorder the internal schema representation based on which columns are required by the query, such that the RDD only loads columns of interest.

Realizing horizontal pruning is more challenging. This holds true especially for handling non-convex and potentially overlapping predicates, as we will see in Section~\ref{ssec:non_convex_query_processing}. For now, let us focus on handling convex predicates. \system{} supports horizontal pruning for arbitrary numeric comparisons as well as SQL's \smalltt{IN} and \smalltt{NOT IN} operators. Further, predicates can be connected arbitrarily via \smalltt{AND} and \smalltt{OR} and be negated via \smalltt{NOT}.
We utilize the subarray lookup provided by the NetCDF API to realize filtering at data-source level.
However, as we have discussed in Section~\ref{ssec:netcdf_api}, a subarray lookup cannot be performed by passing the actual dimension values (as formulated in the query), but requires passing a multidimensional origin and extent. 
Consequently, we have to \textit{translate} all query predicates into the corresponding positional tuples of origin and extent. 
Note that such a translation could be implemented in form of an auxiliary index structure, as conceptually done by ClimateSpark~\cite{lit:climatespark}. However, this would (a)~impose additional storage overhead and (b)~require an expensive preprocessing pass that we want to avoid. Instead, we apply an on-the-fly approach \textit{during} query processing, as we will see in the following.

\subsection{Global and Local Query Rewriting}
\label{sssec:convex_workflow}

When $Q_1$ is submitted, \system{} first detects that the query filters on the two dimensions \smalltt{time} and \smalltt{lat}. As \smalltt{time} is a dimension that spans multiple files, whereas \smalltt{lat} is not, the two dimensions must be treated differently. Also, the \smalltt{lon} dimension, along which $Q_1$ does not filter, must still be considered, as all dimensions are required to perform the subarray lookup.

\system{} expresses the translation of the query predicates to the corresponding origin/extent representation by \textit{rewriting} the query in multiple steps. This rewriting happens on-the-fly during query processing and can be separated into two phases: \textit{global} rewriting and \textit{local} rewriting. Let us see how it works:

The global rewriting of~$Q_1$ starts at the first file of the dataset. For each dimension that does \textit{not} span multiple files, we rewrite all corresponding query predicates in this phase. 
This is the case for the \smalltt{lat} dimension and the \smalltt{lon} dimension. 
Since $Q_1$ filters along the \smalltt{lat} dimension, \system{} reads the array representation of \smalltt{lat} from the file and identifies the positional range that is relevant for our query via binary search. 
As the \smalltt{lat} dimension contains values from $90\degree$ to $-90.0\degree$ in steps of $1/4\degree$, a binary search of the predicate \smalltt{lat~>~20.2} will result in a rewriting of the predicate into \smalltt{latPos~<=~279}. Note that the change in direction of the comparison operation as well as the correct boundaries will be ensured automatically in the process. Similarly, the predicate \smalltt{lat~<~60.5} will be rewritten into \smalltt{latPos~>=~119}. 
As $Q_1$ does not filter along the \smalltt{lon} dimension, \system{} will generate a non-selective predicate in the global rewriting phase. As the dimension has $1440$~entries, we create \smalltt{lonPos~>=~0 AND lonPos~<=~1439}.

Overall, this results in the globally rewritten query $Q_1^{g}$, as shown in Listing~\ref{listing:q1_rewritten1}. 
Note that the core property of the globally rewritten query is that it is valid with respect to the global coordinate system spanning across all files of the dataset.

\noindent 
\begin{minipage}{\linewidth}
\begin{lstlisting}[style=sql, caption={Globally rewritten query~$Q_1^{g}$. Predicates are valid with respect to the global coordinate system of the dataset.}, label={listing:q1_rewritten1}]
SELECT time, lat, lon, sp
FROM era_b
WHERE time > '2017-01-01 00:15:00' AND 
        latPos >= 119 AND latPos <= 279 AND
        lonPos >= 0 AND lonPos <= 1439
\end{lstlisting}
\end{minipage}

%

Note that the \smalltt{time} dimension is not globally rewritten in $Q_1^g$. As \smalltt{time} spans over multiple files of our dataset, a global rewriting of the query predicate \smalltt{time~>~'2017-01-01~00:15:00'} would not yield correct results.
Instead, \system{} performs a local rewriting of the \smalltt{time} predicate in $Q_1^g$, which must happen \textit{individually per file}. Precisely, when processing file~$f$, we read the \smalltt{time} dimension array of $f$ and and identify which entries satisfy our predicate \smalltt{time~>~'2017-01-01~00:15:00'}. In the case of ERA5, the \smalltt{time} dimension of each file only contains one entry (at position~$0$). Thus, if the entry satisfies the predicate, we rewrite the query $Q_1^g$ into $Q_1^{l_{\text{sat}}}$, as shown in Listing~\ref{listing:q1_rewritten2_qual}. If the entry does not satisfy the predicate, we rewrite the query into $Q_1^{l_{\text{unsat}}}$, as shown in Listing~\ref{listing:q1_rewritten2_noqual}. 

\noindent 
\begin{minipage}{\linewidth}
\begin{lstlisting}[style=sql, caption={Locally rewritten query~$Q_1^{l_{\text{sat}}}$, which will be executed on a file that \textbf{satisfies} the \smalltt{time}-predicate.}, label={listing:q1_rewritten2_qual}]
SELECT time, lat, lon, sp
FROM era_b_f -- operates on a specific file f
WHERE timePos >= 0 AND timePos <= 0 AND -- all
        latPos >= 119 AND latPos <= 279 AND
        lonPos >= 0 AND lonPos <= 1439
\end{lstlisting}
\end{minipage}

\noindent 
\begin{minipage}{\linewidth}
\begin{lstlisting}[style=sql, caption={Locally rewritten query~$Q_1^{l_{\text{unsat}}}$, which will be executed on a file that \textbf{does not satisfy} the \smalltt{time}-predicate.}, label={listing:q1_rewritten2_noqual}]
SELECT time, lat, lon, sp
FROM era_b_f -- operates on a specific file f
WHERE timePos > 0 AND timePos <= 0 AND -- empty set
        latPos >= 119 AND latPos <= 279 AND
        lonPos >= 0 AND lonPos <= 1439
\end{lstlisting}
\end{minipage}

After a local query has benn generated for a file, \system{} starts the retrieval of the projected dimensions and variables. While \system{} retrieves the projected dimensions individually via a single-dimensional lookup, 
it performs a final translation from the predicates of the local query to the origin/extent representation to retrieve the variables. For example, to retrieve the project variable~\smalltt{sp}, $Q_1^{l{\text{sat}}}$ is translated to origin
$o=(\smalltt{time}=0, \smalltt{lon}=0, \smalltt{lat}=119)$
and extent
$e=(\smalltt{time}=0, \smalltt{lon}=1439, \smalltt{lat}=160)$, then the subarray lookup is performed. Note that at this stage, \system{} also detects that the \smalltt{time}-predicate of~$Q_1^{l_{\text{unsat}}}$ results in an empty set. Thus, it eliminates this query at this stage entirely.

As previously described, \system{} materializes the data in a convenient row-wise representation after loading. 
Note that the result of a single subarray lookup can potentially yield enough entries to exceed the available main memory capacity of the worker node. Thus, before performing the lookup, we split a single lookup into multiple smaller ones if necessary. When splitting, we avoid splitting the \textit{fastest-varying dimension}, which is stored continuously in memory, to preserve the sequential read performance.

\section{Non-Convex Query Processing \& Optimization}
\label{ssec:non_convex_query_processing}

So far, we have the discussed how to efficiently process convex predicates, such as shown in query~$Q_1$. Again, convex means that all predicates of the query result in the selection and consequently the loading of a \textit{single} block per file.
However, various applications in the Earth science community often require the formulation of non-convex predicates and can therefore require multiple blocks per file. For example, certain atmospheric analyses require the area of interest to be limited to regions covered by water, such as oceans. Also, certain areas are also often excluded from the analysis, creating "holes" in the region of interest. Of course, we could simply load a larger convex block enclosing all blocks described by the non-convex predicates and then filter the result afterwards. This is the only option in ClimateSpark, which cannot automatically push down such a non-convex predicate to the data source. However, this could result in loading significant amounts of unnecessary data. Instead, we want to ensure that only the data of interest is actually loaded from the data source. 

\subsection{Overview}

To do so, \system{} again applies a set of rewriting rules to the non-convex predicate in order to generate a set of convex predicates. Each generated convex block can then be loaded via a single subarray lookup, as done before. While doing so, two circumstances must be respected:
\begin{enumerate} 
\item \label{enum:overlap} The generated convex blocks potentially \textit{overlap arbitrarily} across one or multiple dimensions.
\item \label{enum:many} We might generate a \textit{large number of convex blocks} for a single query.
\end{enumerate}
Regarding~\ref{enum:overlap}), we have to ensure \textit{correctness} in case of overlapping blocks, i.e. we must not load data within overlapping regions multiple times. 
Regarding~\ref{enum:many}), we may have to handle a large number of blocks during overlap detection and subsequent lookups. Thus, we require efficient overlap detection strategies as well as reduce the number of generated blocks again before the actual subarray lookups are performed. 
In the following, we will describe how we deal with the aforementioned challenges.

%


In short, we first rewrite the \smalltt{WHERE} clause, which consists of potentially many non-convex query predicates, into \textit{disjunctive normal form}, i.e. a disjunction of conjunctive clauses. This automatically generates a set of convex predicates, which individually represent blocks to load. 
Section~\ref{ssec:non_convex_to_convex} discusses the details of this step.
Note that this conversion could theoretically inflate the predicate exponentially, but we do not expect this to happen in practice.

Next, we globally rewrite the predicate according to the steps described in Section~\ref{ssec:convex_query_processing}. After rewriting the query globally, we ensure that no region of the dataset is loaded more than once to ensure correctness. In Section~\ref{ssec:overlap}, we propose multiple strategies to do so. The final transformation into local queries is performed afterwards.

\subsection{Non-Convex to Convex Predicate Rewriting}
\label{ssec:non_convex_to_convex}

Let us consider a concrete example. Query~$Q_2$, shown in Listing~\ref{listing:q2}, selects a certain surface region based on the \smalltt{lon} and \smalltt{lat} dimension. As we can see, the selection \textit{excludes} a certain subregion.
\begin{minipage}{\linewidth}
\begin{lstlisting}[style=sql, caption={Example query~$Q_2$ (non-convex predicates).}, label={listing:q2}]
SELECT time, lat, lon, sp
FROM era_b
WHERE lon >= 90.0 AND
 NOT (lat == 0.0 AND lon >= 163.0 AND lon <= 163.75)
\end{lstlisting}
\end{minipage}

\noindent In the first step, the \smalltt{WHERE} clause is automatically rewritten into disjunctive normal form (DNF). Listing~\ref{listing:q2_where_rewriting} shows the individual steps that are carried out. Initially, we eliminate any present negations; this changes the comparison and logical operators. Then, we express all equalities by the corresponding inequalities. Finally, we expand the clause into DNF. 

\begin{lstlisting}[style=sql, caption={Individual rewriting-steps of query~$Q_2$ into DNF.}, label={listing:q2_where_rewriting}]
-- original predicate
lon >= 90.0 AND 
NOT (lat == 0.0 AND lon >= 163.0 AND lon <= 163.75)
-- eliminate negation
lon >= 90.0 AND 
(lat != 0.0 OR lon < 163.0 OR lon > 163.75)
-- rewrite equalities into inequalities
lon >= 90.0 AND 
(lat < 0.0 OR lat > 0.0 OR 
  lon < 163.0 OR lon > 163.75)
-- disjunctive normal form
(lon >= 90.0 AND lat < 0.0) OR 
(lon >= 90.0 AND lat > 0.0) OR
(lon >= 90.0 AND lon < 163.0) OR 
(lon >= 90.0 AND lon > 163.75)
\end{lstlisting}

With the DNF at hand, we are ready to generate the globally rewritten query, shown in Listing~\ref{listing:q2_global}. The \smalltt{WHERE}-clause contains the predicates in DNF, where each predicate which operates on a dimension that does not span across files has been translated into the corresponding positional boundaries. Again, this changes the comparison operator for the \smalltt{lat} dimension. Further, in this step, we have added the non-selective predicate for the missing \smalltt{time} dimension. As \smalltt{time} spans across multiple files, we formulate this predicate with respect to the global coordinate system.

\noindent 
\begin{minipage}{\linewidth}
\begin{lstlisting}[style=sql, caption={\smalltt{WHERE}-clause of globally rewritten query~$Q_2^{g}$.}, label={listing:q2_global}]
WHERE (time >= minTime AND time <= maxTime AND
         lonPos >= 361 AND lonPos <= 1439 AND
         latPos >= 362 AND latPos <= 721)
         OR
         (time >= minTime AND time <= maxTime AND
         lonPos >= 361 AND lonPos <= 1439 AND
         latPos >= 0 AND latPos <= 360) 
         OR
         (time >= minTime AND time <= maxTime AND
         lonPos >= 361 AND lonPos <= 652 AND
         latPos >= 0 AND latPos <= 721) 
         OR
         (time >= minTime AND time <= maxTime AND
         lonPos>= 657 AND lonPos <= 1439 AND
         latPos >= 0 AND latPos <= 721) 
\end{lstlisting}
\end{minipage}

\noindent Note that in the last conjunctive clause, \system{} translated 
\mbox{\smalltt{(lon >= 90.0 AND lon > 163.75)}} into 
\mbox{\smalltt{lon > 163.75}} and subsequently into
\smalltt{(lonPos >= 657 AND lonPos <= 1439)}, as the predicate \smalltt{lon > 163.75} is more selective than \smalltt{lon >= 90.0}.
In summary, we get four conjunctive clauses, where each clause selects a block.

Note that the same predicate transformation has been performed by Wang et al.~\cite{lit:rewriting} to provide a query interface for HDF5~\cite{list:hdf5} files, which are structurally similar to NetCDF-4 files.
However, Wang et al. do not mention the case where multiple conjunctive clauses describe overlapping regions, which can lead to incorrect results as we will discuss in the next section.



\subsection{Overlap Detection and Elimination}
\label{ssec:overlap}

As we see from Listing~\ref{listing:q2_global}, the resulting blocks overlap across multiple dimensions. If we translated each block into a corresponding subarray lookup, we would generate incorrect results. Thus, we need to transform the generated blocks such that no overlap occurs anymore. To do so, we propose two different strategies, which we will discuss in the following. 

Let $n$ be the number of blocks and $d$ be the number of dimensions.
A single block corresponds to a lower and an upper bound along each dimension,
so there are at most $2n$ distinct bounds along each dimension, and at most $2n - 1$ intervals between adjacent bounds.

\subsubsection{Naive}
\label{sssec:strategy1}
In the naive strategy, we iterate over all blocks and split each block at each distinct interval boundary along all dimensions. This generates at most $(2n - 1)^d$~sub-blocks per block. The resulting blocks either overlap entirely or not at all. Thus, we remove all blocks that overlap next. Finally, we merge adjacent blocks into larger ones to keep the number of resulting blocks as small as possible. To improve performance when loading blocks from disk later on, we merge along the fastest-varying dimension of the dataset first, resulting in blocks having the largest possible extent along this dimension. 

\begin{figure}[h!]
\vspace*{-0.3cm}
\subfloat[Split of blocks into either fully overlapping or non-overlapping sub-blocks.]{
\includegraphics[page=1, width=\columnwidth, trim={0 11cm 0 0}, clip]{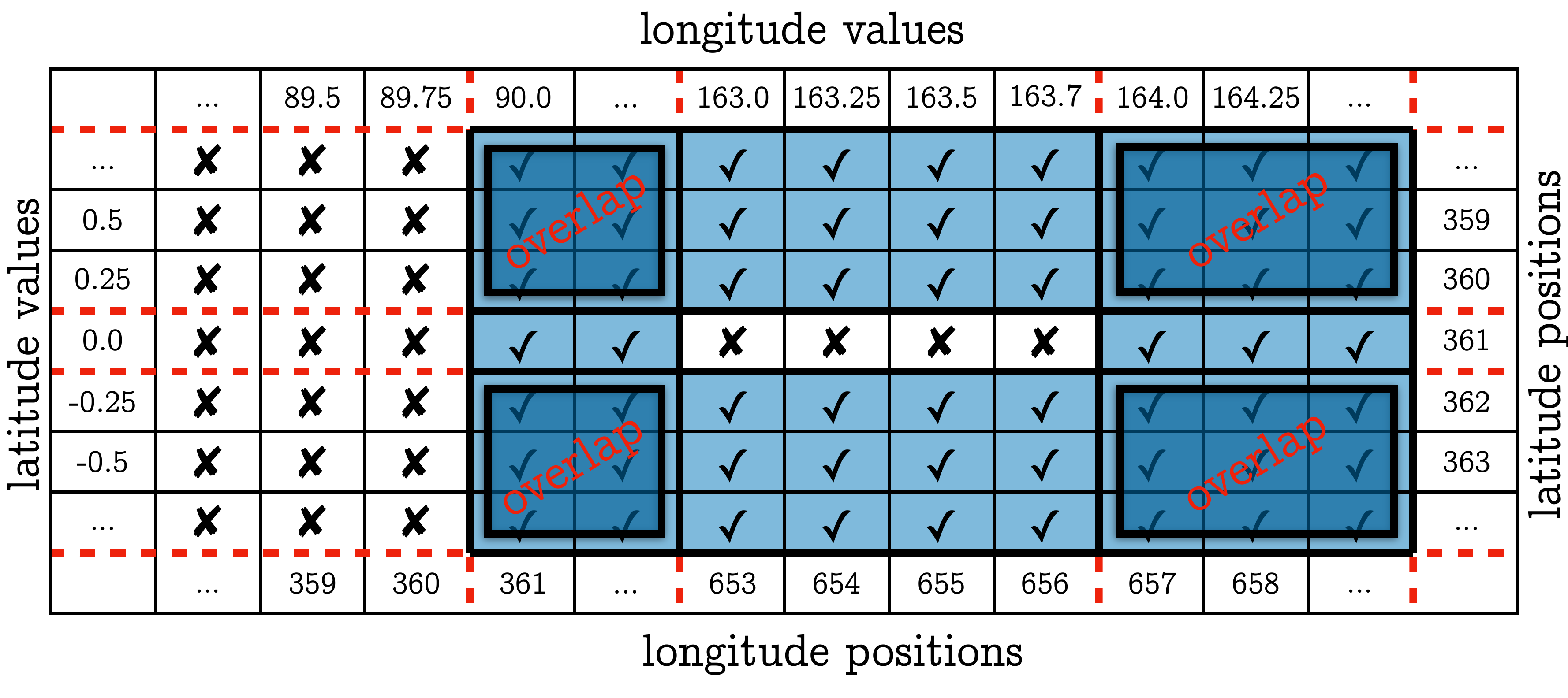}
\label{fig:strategy1:split}
}\quad
\subfloat[After elemination of overlapping sub-blocks, merge of adjacent sub-blocks.]{
\includegraphics[page=2, width=\columnwidth, trim={0 11cm 0 0}, clip]{non_convex_small.pdf}
\label{fig:strategy1:merge}
}
\caption{Overlap elimination using the naive strategy.}
\vspace*{-0.4cm}
\label{fig:strategy1}
\end{figure}

Let us see how the naive strategy would eliminate the overlaps of the $n=4$ blocks of query~$Q_2^g$. We have $4$~distinct bounds along both the \smalltt{lat} and the \smalltt{lon} dimension, with $3$~intervals between adjacent bounds for each dimension. The \smalltt{time} dimension is non-selective and therefore contains only one interval. Now, we split the blocks along each dimension, as shown in Figure~\ref{fig:strategy1:split}. This results in $12$~independent sub-blocks. We then remove the 4 redundant blocks by inserting all blocks into a \smalltt{HashSet}, resulting in $8$~remaining sub-blocks in total. Finally, as shown in Figure~\ref{fig:strategy1:merge}, we can merge the $6$~adjacent sub-blocks into $2$~larger ones along the fastest-varying dimension. Overall, we get $4$~non-overlapping blocks.

\subsubsection{Optimized}
\label{sssec:strategy2}

The naive strategy has two limitations: We might generate a large number of sub-blocks in the process of splitting. For only $n=4$ and $d=3$, we generate up to $(2n - 1)^d = 343$~sub-blocks per block in the worst case. Thus, the number of generated sub-blocks might (a)~exceed the available memory capacity of the driver node and (b)~render the overlap detection and subsequent merge expensive. In the following, we thus propose an optimized version to address these issues. 
\begin{lstlisting}[style=pseudo, caption={Pseudo-code of the optimized strategy.}, label={listing:strategy2}, xleftmargin=4.0ex]
findCover(blocks, dim=0, lastDim):
   // find 1D interval cover if applicable
   if dim == lastDim:
      return findIntervalCover(blocks, lastDim)
   cover <- []
   boundaries <- [for b in blocks: b.start[dim]] 
                   + [for b in blocks: b.end[dim]]
   boundaries.sort()
   boundaries.removeDuplicates()
   // split the blocks along adjacent boundaries
   for i in range(boundaries.size - 1):
      start, end = boundaries[i], boundaries[i + 1]
      active <- blocks that overlap with [start, end]
      sliced <- copy(active)
      for b in sliced:
         b.start[dim], b.end[dim] = start, end
      // recursively find a cover for this subset
      subCover = findCover(sliced, dim + 1, lastDim)
      // merge blocks before the next iteration
      cover <- mergeAlignedBlocks(subCover, cover)
   return cover
\end{lstlisting}
In the optimized strategy, we first sort the distinct intervals along a specific dimension. Then, we process these intervals in sorted order one at a time: For each interval, we split only along the interval boundaries and try to merge the newly obtained sub-blocks with those obtained in the previous iteration. Listing~\ref{listing:strategy2} shows the workflow in pseudo-code. 
By applying this strategy recursively along each dimension, the intermediate result is kept as small as possible at all times, addressing problem~(a) of the naive strategy. Further, the set of sub-blocks on which we have to detect overlaps and perform the merging remains small, effectively addressing problem~(b) of the naive strategy.  
Further, we process the fastest-varying dimension in the innermost recursion step. Thus, we still generate blocks that allow fast sequential access. Overall, this strategy produces the same result as the naive strategy, albeit more efficiently, as we will evaluate in Section~\ref{sec:experiments}.

\begin{figure}[h!]
\vspace*{-0.3cm}
\subfloat[Split by the first interval of \scriptsizett{lat} dimension to eliminate overlapping blocks.]{
\includegraphics[page=3, width=\columnwidth, trim={0 11cm 0 0}, clip]{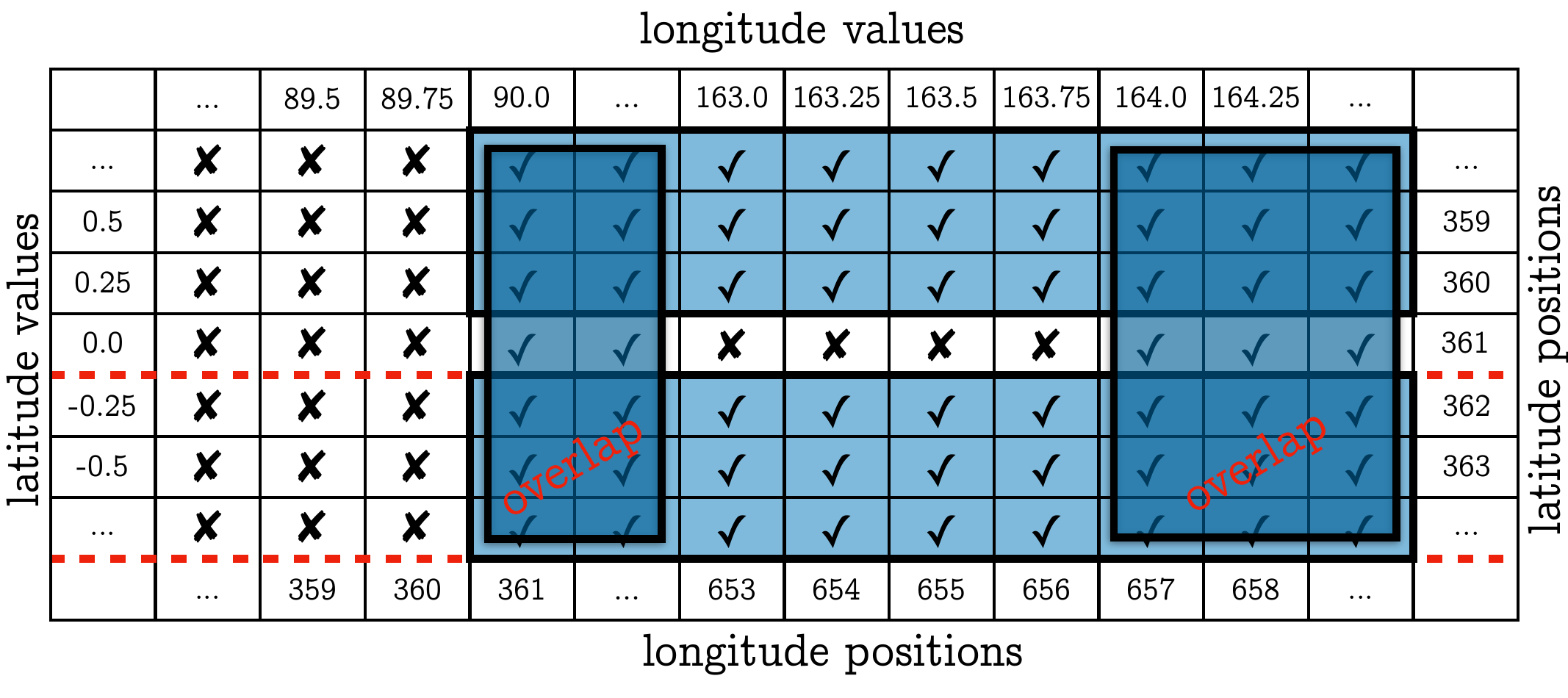}
\label{fig:strategy2:interval1}
}\quad
\subfloat[Split by the second interval of \scriptsizett{lat} dimension. This time, no overlap must be eliminated. Try to merge the sub-blocks of the current iteration with the sub-blocks of the previous iteration, which is not possible here.]{
\includegraphics[page=4, width=\columnwidth, trim={0 11cm 0 0}, clip]{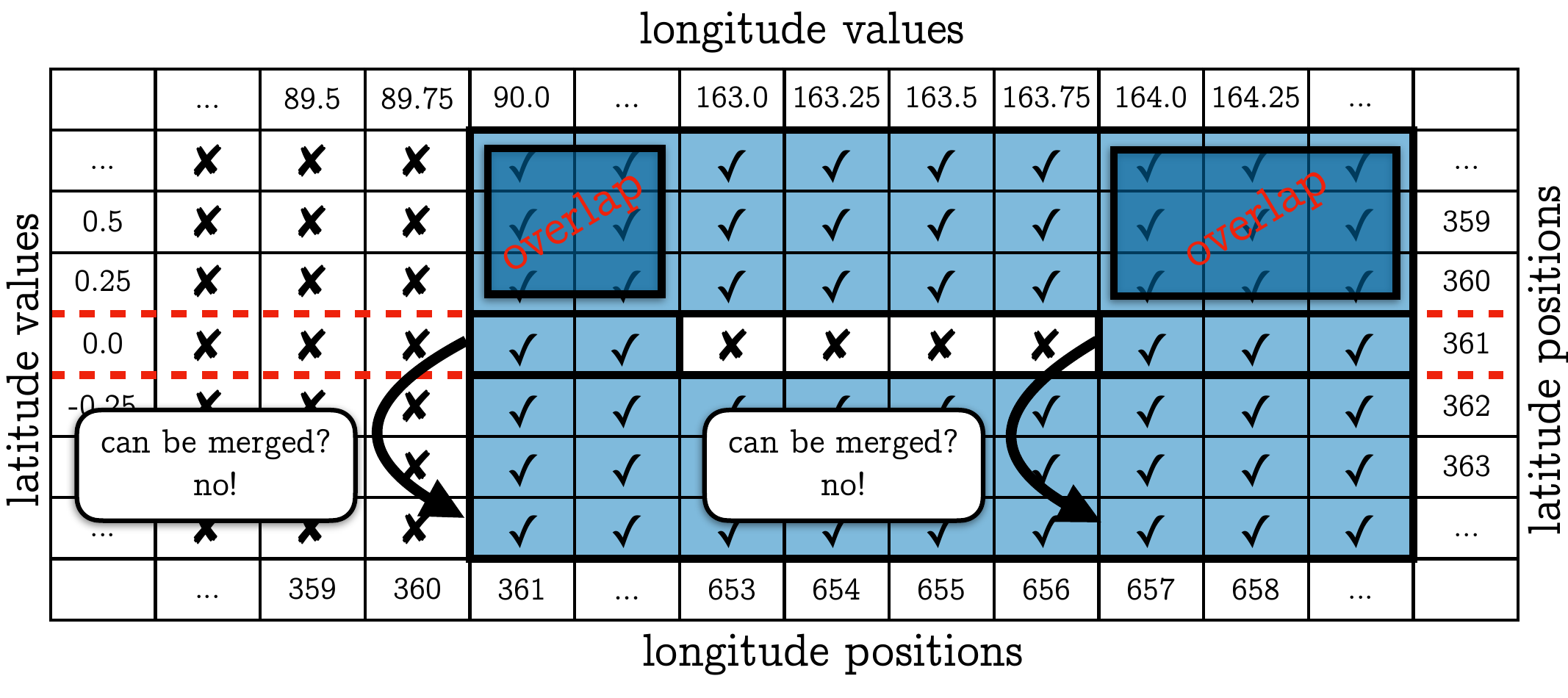}
\label{fig:strategy2:interval2}
}
\caption{Overlap elimination using the optimized strategy.}
\vspace*{-0.4cm}
\label{fig:strategy2}
\end{figure}

In Figure~\ref{fig:strategy2}, we show how the optimized strategy processes two intervals along the \smalltt{lat} dimension. In Figure~\ref{fig:strategy2:interval1}, which shows the handling of the first interval, we split $2$~blocks along the boundary between latitude~$-0.25$ and $0.0$. Afterwards, we detect that the generated sub-blocks overlap with the third block of this area, and consequently, eliminate both. Next, in Figure~\ref{fig:strategy2:interval2}, which shows the handling of the second interval, we again split $2$~blocks, this time at the boundary between latitude~$0.0$ and $0.25$. Again, we check for an overlap of the newly created sub-blocks, however, there are no overlaps this time. As a final step of the round, we check whether the newly created sub-blocks can be merged directly with the sub-blocks created when handling the previous interval. In this case, this is not possible and thus, this interval is fully processed. The procedure repeats until all intervals have been processed. 

\section{Join Optimization via Envelopes}
\label{sec:envelopes}

In the context of Earth science, analytical tasks often require the joining of multiple independent datasets or subsets of those. For example, the ERA5 dataset is split into one subset consisting of surface data and into another model-level subset with an additional vertical dimension -- often, both are required together and must be joined consequently. As such, we want to enable automatic and early horizontal pruning of data in the presence such joins as well. 
Precisely, we target the common situation where dataset~$R$ is more selective than dataset~$S$ with respect to the dimension on which the join is performed. In this situation, we want to avoid that~$S$ is fully loaded since many entries of~$S$ will not find a join partner in $R$. Of course, we could require the user to add a corresponding filtering condition on $S$ to the query. However, this requires a manual adjustment according to the characteristics of the underlying datasets and thus stands in conflict with \system{}'s principle of applying \textit{automatic} optimizations wherever possible. 

Instead, \system{} support the generation and automatic usage of so-called \textit{envelopes}. Envelopes are a set of tight lower and upper bounds for each file of a dataset. The user can conveniently request to generate envelopes for a specific dataset and a set of dimensions when registering the dataset in SparkSQL. By calling our \smalltt{envelope()}~function on the DataFrame, the auxiliary envelope information will be generated and used during the processing of any query that operates on the enveloped dataset:
\noindent
\begin{minipage}{\linewidth}
\begin{lstlisting}[style=sql, label={listing:envelopes}]
envelope(ERA_Sur, "time", "longitude", "latitude")
   .createOrReplaceTempView("Era_Sur")
\end{lstlisting}
\end{minipage}

The challenge lies in communicating to SparkSQL that only such data of~$S$ should be loaded that lies within the envelopes of $R$. While SparkSQL implements this form of invariant propagation internally\cite{lit:invariants}, unfortunately, it does not expose access to this mechanism in its public API.
To overcome this limitation, we gather all envelopes into a single filter operation and apply this filter on the DataFrame of~$R$ during the call of \smalltt{envelope()}. Note that this filter does not actually remove data in $R$ because each record of $R$ is contained within the envelope. However, this filter information of $R$ will be picked up by the optimizer of SparkSQL and pushed down to the other side of the join, namely $S$. On $S$, it results in horizontal pruning at the data source, as described before. 

Of course, this filter operation also introduces execution overhead, as SparkSQL has to evaluate a complex auxiliary predicate for every record of both $R$~and~$S$, in addition to the cost of computing the envelope. Still, as we will see in Section~\ref{sec:experiments}, the overhead is compensated by having to load significantly less data from disk. 

\section{Experimental Evaluation}
\label{sec:experiments}


\subsection{Setup and Datasets}

All tests were conducted on a SparkSQL~3.0.0 cluster with $8$~executors per node with $8$~cores each. If not stated otherwise, we use $5$~nodes in total. The cluster is set up on top of MOGON~\cite{lit:mogon}, the HPC infrastructure at Mainz University, where each node consists of two 16-core Xeon Gold 6130 CPUs paired with 177 GB of RAM and an OmniPath interconnect. The dataset is stored on a Lustre filesystem~\cite{lit:lustre} instead of HDFS, so \system{} has to ignore data locality for files.
All workers are configured to use a node-local SSD for temporary storage. Unless mentioned otherwise, each experiment was run exactly once to avoid caching artifacts.

We use the ERA5 dataset~\cite{lit:ecmwf-era5,lit:era5} for our evaluation, where we use the following three subsets:
\textbf{ERA5\_Sur}:~The 3D surface subset ($38$MB/file), which stores, among others, snow cover and surface pressure, as well as wind speeds and temperature just above the surface.
\textbf{ERA5\_Pre}:~The 3D precipitation subset, which contains rain and snowfall data ($26$MB/file).
\textbf{ERA5\_Mod}:~The 4D model-level subset, which includes an additional vertical dimension that ranges from the surface to the top of the atmosphere ($2.2$GB/file). Variables include specific humidity, but also wind speeds and temperature for each model level separately.
Each subset is stored in a separate set of files. 
All three subsets use the $1/4\degree$ by $1/4\degree$ surface grid mentioned before, with a temporal resolution of one hour.

The majority of our benchmark queries is inspired by real-world analysis tasks from atmospheric physics.
In some of the queries, we mention a quantity called \textit{relative humidity with respect to ice} (RHI)~\cite{lit:rhi}, which indicates the amount of water vapor with respect to the stable phase ice.
This quantity is calculated through a formula\cite{lit:rhi-formula} that depends on variables from both the surface and model-level subsets.


\subsection{Experiment: Effect of Pruning}

\begin{figure}[h!]
\vspace*{-0.7cm}
\subfloat[Effect of horizontal pruning.]{
\includegraphics[width=43mm]{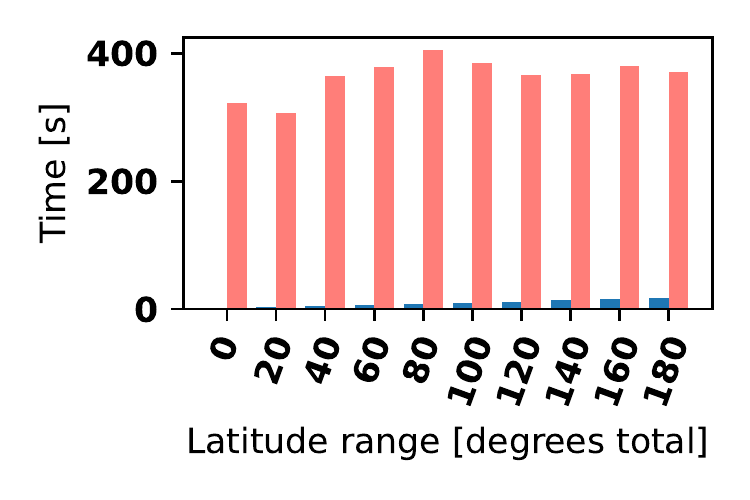}
\label{fig:opt:exp1}
}
\subfloat[Effect of vertical pruning.]{
\includegraphics[width=43mm]{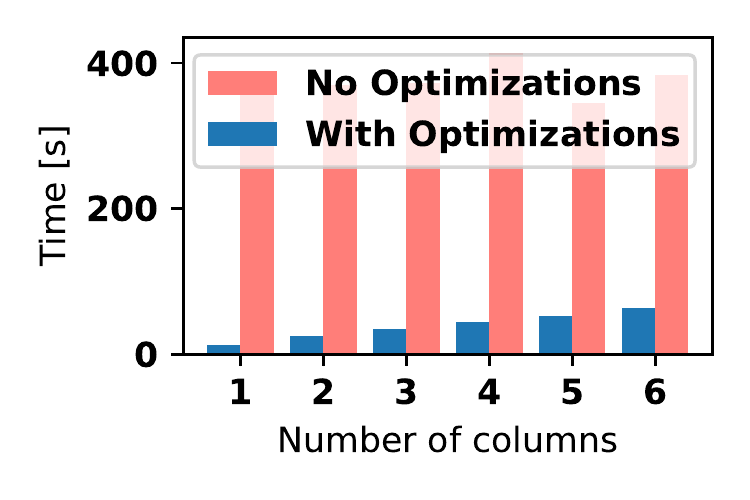}
\label{fig:opt:exp2}
}
\vspace*{-0.1cm}
\caption{Evaluation of Pruning Optimizations.}
\vspace*{-0.3cm}
\label{fig:opt}
\end{figure}

We start with two micro-benchmarks to evaluate the effect of vertical and horizontal pruning on the ERA5\_Mod subset.
In Figure~\ref{fig:opt:exp1}, we fire a query that filters a single model-level file along the \smalltt{lat} dimension and then calculates the mean of the variable~\smalltt{t} (temperature). We vary the selected range of the \smalltt{lat} dimension from $0^\degree$ to $180\degree$.
In Figure~\ref{fig:opt:exp2}, we fire a query that does not perform any filtering of the file, but computes the mean of a certain number of variables, which we vary from~$1$ to $6$ (\smalltt{t, cc, q, o3, u, v}). We evaluate \system{} both with activated and deactivated optimizations.

As we can see in Figure~\ref{fig:opt}, the execution time is significantly increased and does not reflect the query's selectivity when all optimizations are disabled.
This is because Spark has to fully read every single entry in the file and can only apply the projections and selections after a tuple has been loaded from the data source.
We can also see that the execution time increases linearly with the amount of data loaded if our optimizations are enabled.
In absolute terms, enabling pruning reduces the execution time of the query from well above five minutes to $60$~seconds, and in many cases far below that.
Note that since each file corresponds to a single partition, this experiment does \textit{not} include any form of parallelism, even through there are multiple nodes available.

\subsection{Experiment: Real-world Atmospheric Application}

As \system{} is meant to support atmospheric research, let us evaluate a real-world application from the domain, namely the computation of RHI histograms. 
\begin{wrapfigure}{r}{0.475\columnwidth}
\vspace*{-0.8cm}
\subfloat{
\hspace*{-0.2cm}\includegraphics[width=43mm]{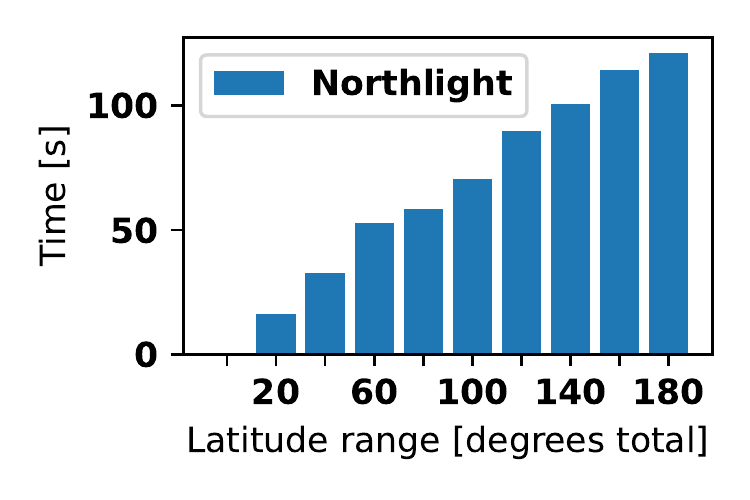}
\label{fig:data-load:lat}
}\\ \vspace*{-0.5cm}
\subfloat{
\hspace*{-0.2cm}\includegraphics[width=43mm]{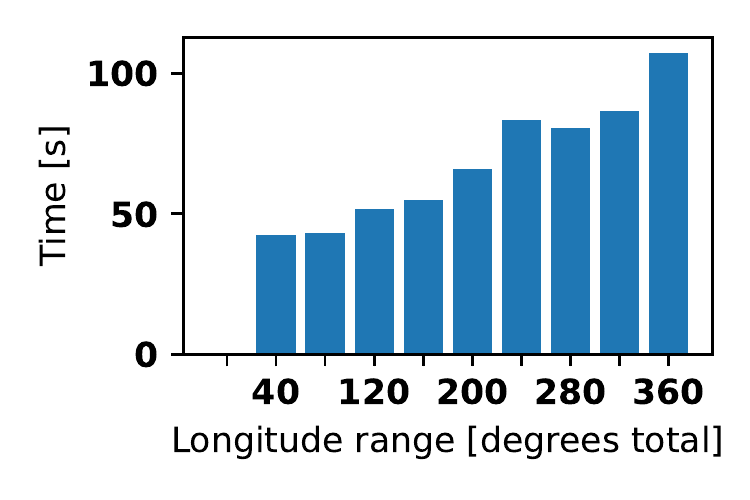}
\label{fig:data-load:lon}
}\\ \vspace*{-0.5cm}
\subfloat{
\hspace*{-0.2cm}\includegraphics[width=43mm]{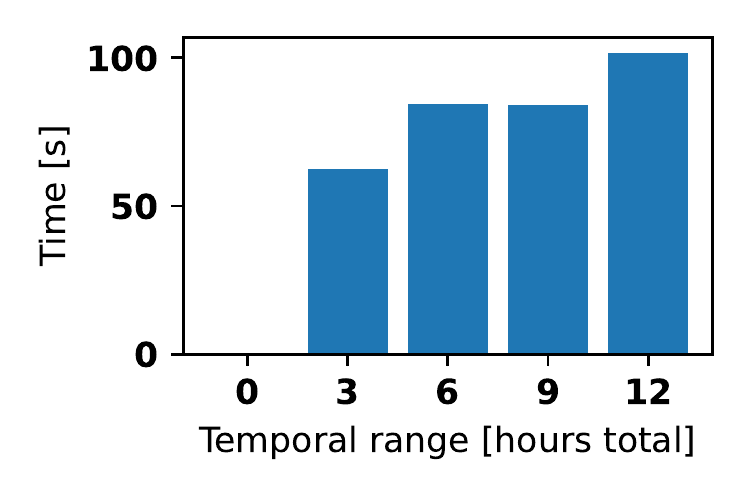}
\label{fig:data-load:time}
}
\caption{RHI histogram.}
\vspace*{-0.2cm}
\label{fig:data-load}
\end{wrapfigure}
We provide one month of data from both ERA5\_Mod and ERA5\_Sur, amounting to a total of $1.6$TB stored across $1488$~files. 
In the experiment, we evaluate queries which join these two subsets along their \smalltt{time}, \smalltt{lat}, and \smalltt{lon} dimensions, and then compute a RHI histogram. We vary the following restrictions: 
In Figure~\ref{fig:data-load:lat}, we restrict the time dimension to a 12-hour window and vary the extent along the \smalltt{lat} dimension. In Figure~\ref{fig:data-load:lon}, we restrict the \smalltt{time} dimension to a 12-hour window and vary the extent along the \smalltt{lon} dimension. In Figure~\ref{fig:data-load:time}, we vary the temporal window between 0 and 12 hours.

As we can see in Figure~\ref{fig:data-load}, the performance of \system{} scales gracefully with the amount of data being selected by the computation under this real-world application. Note that the last bar in each plot corresponds to loading the entire 12-hour window, and results in a consistent time between experiments.


\vspace*{-0.1cm}
\subsection{Experiment: Effect of Envelopes on Join Performance}

Let us now evaluate the effect of generating envelopes to speed up join processing. 
We set up the experiment by cutting out a contiguous six-hour window covering the North Atlantic Ocean from ERA5\_Sur, retaining only the \smalltt{time}, \smalltt{lat}, \smalltt{lon}, and \smalltt{sp} columns, and export it to CSV format using Spark. We call this artificial dataset ERA5\_NAO. Note that ERA5\_NAO is more selective on the join dimensions than the remaining ERA5 subsets. We then join this dataset with $12$~files from ERA5\_Mod and measure the execution time.
The $12$~files are selected to contain all six hours from the artificial dataset, so that each row in ERA5\_NAO has a matching join partner. After performing the join, we count the number of records where the RHI exceeds 100\% and the temperature remains below 243 Kelvin.

\begin{figure}[h!]
\vspace*{-0.1cm}
\includegraphics[width=90mm]{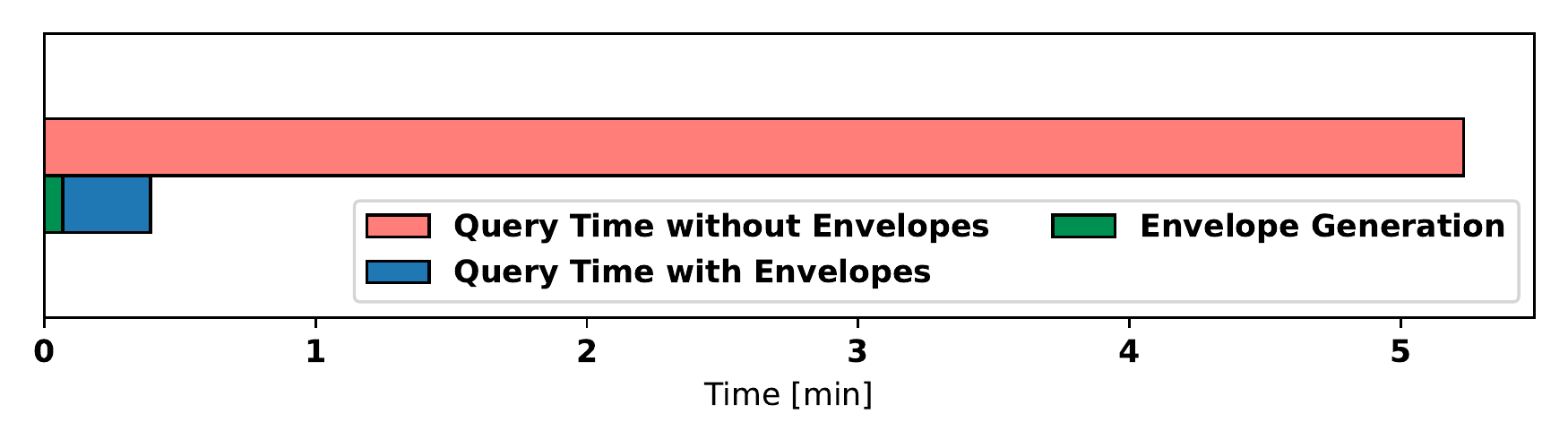}
\vspace*{-0.7cm}
\caption{Effect of Envelopes on Join Processing.}
\vspace*{-0.5cm}
\label{fig:envelopes}
\end{figure}

Figure~\ref{fig:envelopes} shows the runtime of the computation with and without the generation of envelopes on ERA5\_NAO. We can see that the generation of envelopes, which is included in the runtime, reduces the end-to-end processing time by a factor of almost $12$x.  This is due to the reduction in I/O operations for the ERA5\_Mod side of the join.

\subsection{Experiment: \system{} vs ClimateSpark pipeline}

Next, we want to evaluate \system{} against a comparable baseline, namely ClimateSpark~\cite{lit:climatespark}. Unfortunately, the only publicly available version~\cite{lit:climatespark-github} of ClimateSpark depends on a legacy version of Spark and is not compatible with Spark 3.0.0. In addition, the implementation is specifically hard-coded against the MERRA and MERRA2 datasets and would require deep changes to support ERA5. 

\begin{figure}[h!]
\vspace*{-0.7cm}
\subfloat[Maximum wind speed.] {
\includegraphics[width=43mm]{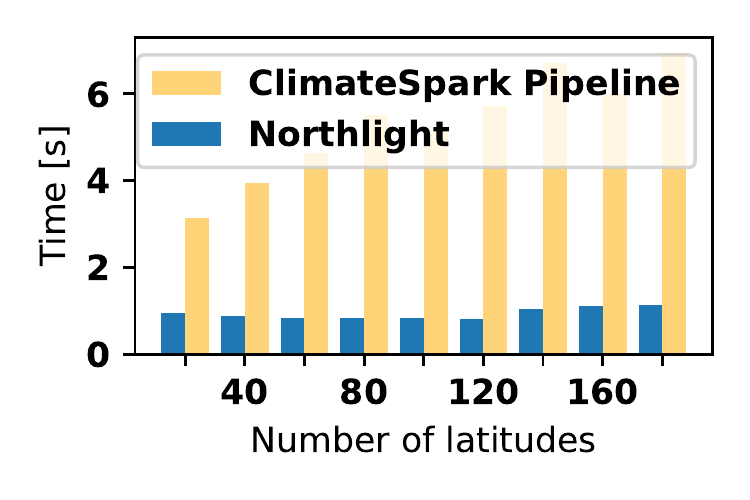}
\label{fig:climatespark:wind}
}
\subfloat[Maximum convective precipitation.] {
\includegraphics[width=43mm]{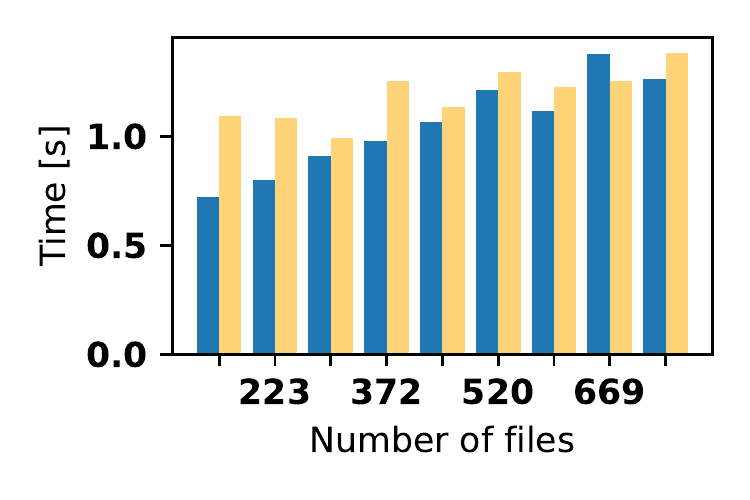}
\label{fig:climatespark:oceans}
}
\caption{\system{} vs ClimateSpark-Pipeline.}
\vspace*{-0.3cm}
\label{fig:climatespark}
\end{figure}

\begin{figure*}[h!]
\newcolumntype{C}{ >{\centering\arraybackslash} m{42mm} }
\begin{tabular}{>{\centering\arraybackslash} m{5mm} @{} C @{} C @{} C @{} C}
& \hspace*{2mm} WL\_aligned & \hspace*{2mm} WL\_misaligned & \hspace*{2mm} WL\_diagonal & \hspace*{2mm} WL\_centered \\
\rotatebox{90}{\hspace*{4mm}d = 1} & \includegraphics[width=40mm]{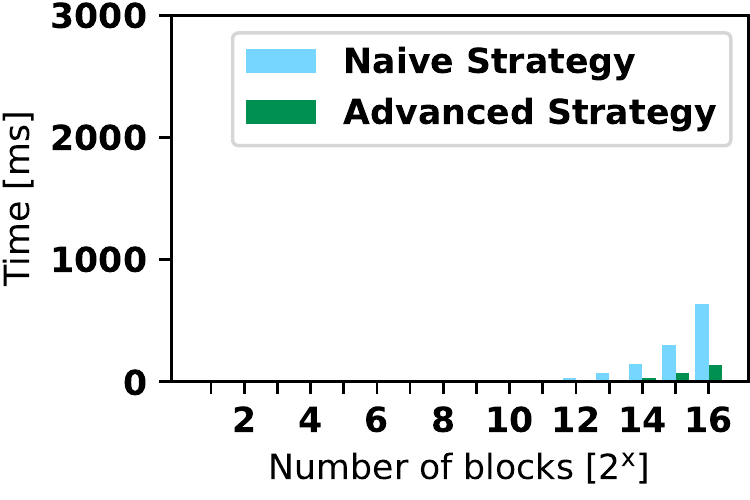} & \includegraphics[width=40mm]{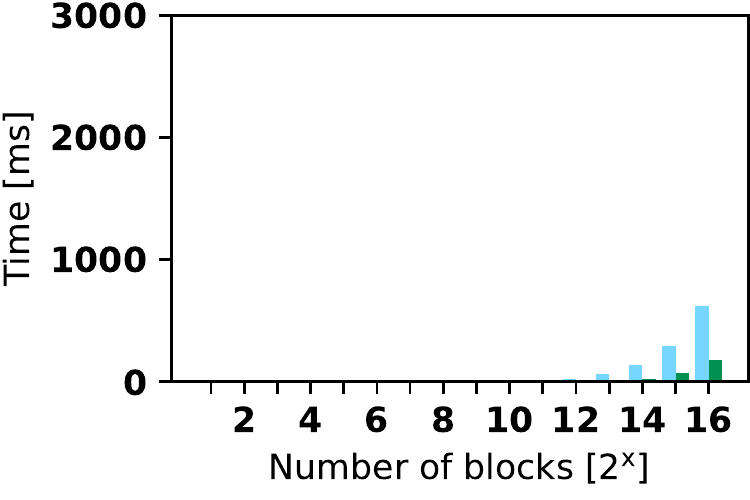} & \includegraphics[width=40mm]{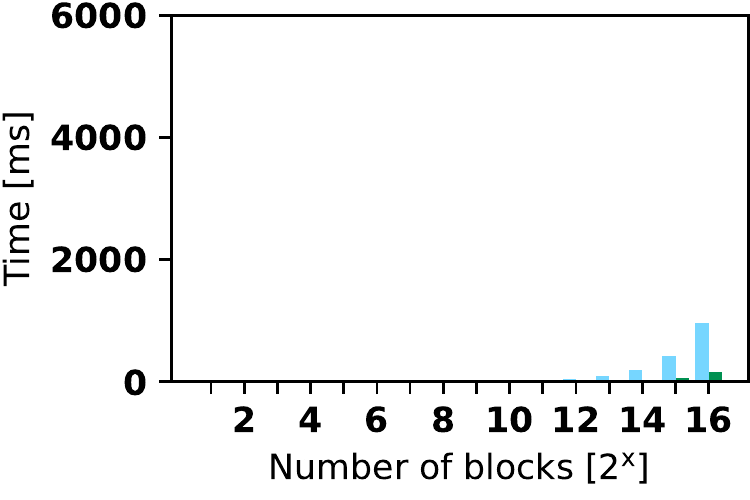} & \includegraphics[width=40mm]{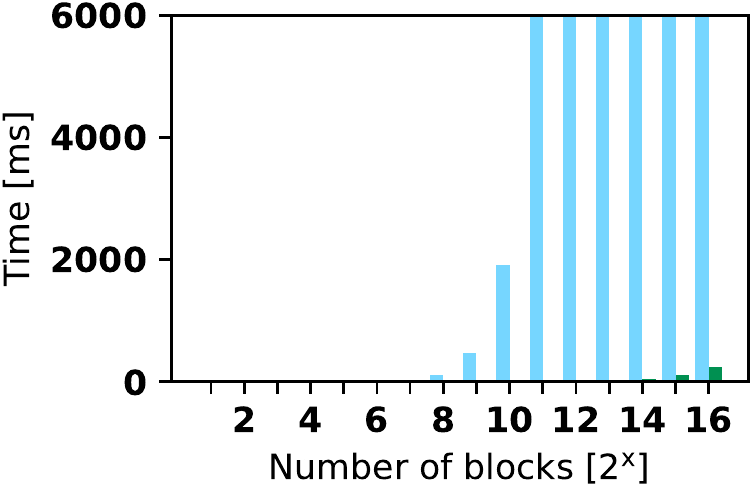} \\
\rotatebox{90}{\hspace*{4mm}d = 2} & \includegraphics[width=40mm]{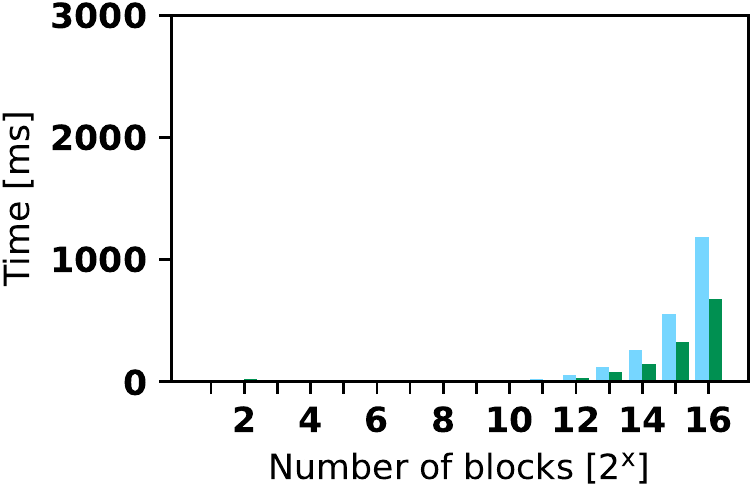} & \includegraphics[width=40mm]{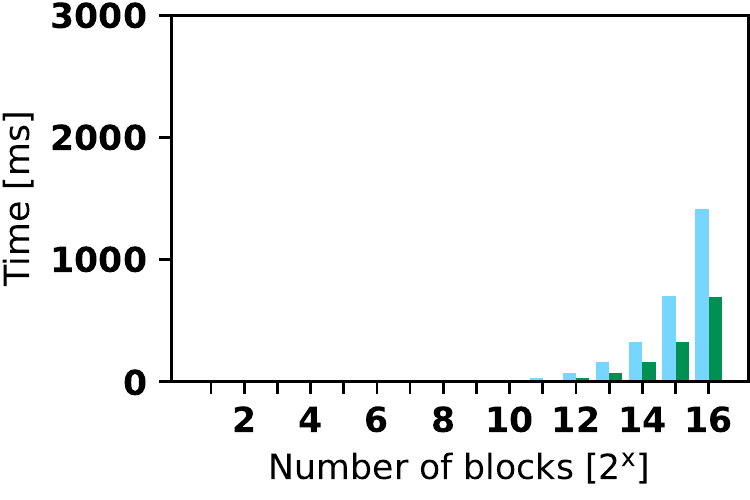} & \includegraphics[width=40mm]{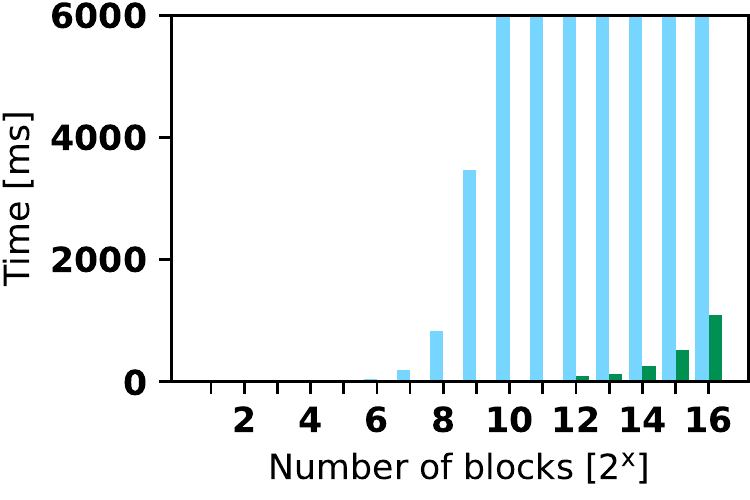} & \includegraphics[width=40mm]{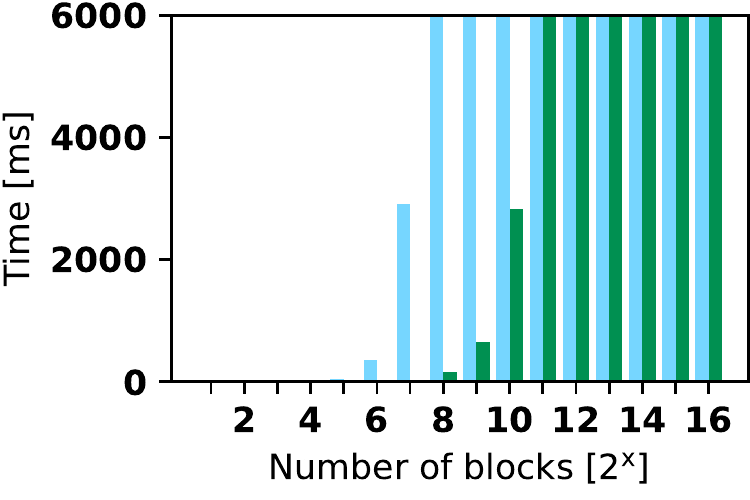} \\
\rotatebox{90}{\hspace*{4mm}d = 4} & \includegraphics[width=40mm]{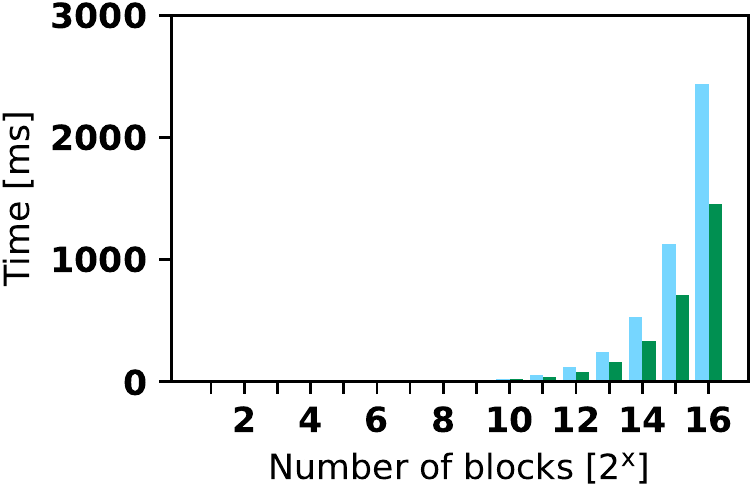} & \includegraphics[width=40mm]{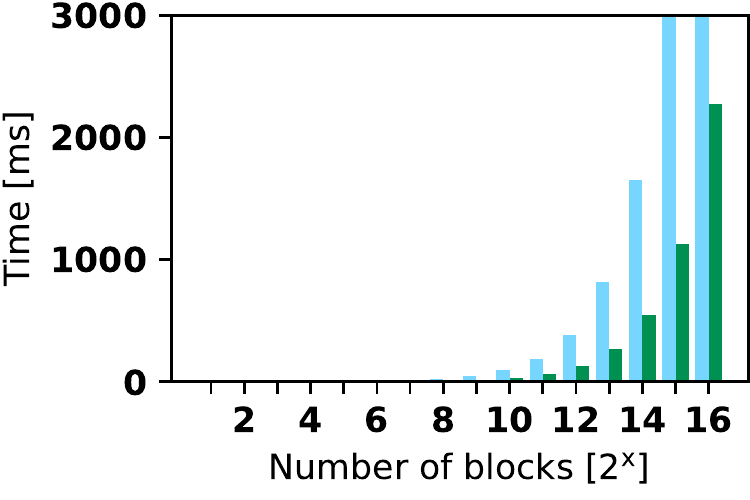} & \includegraphics[width=40mm]{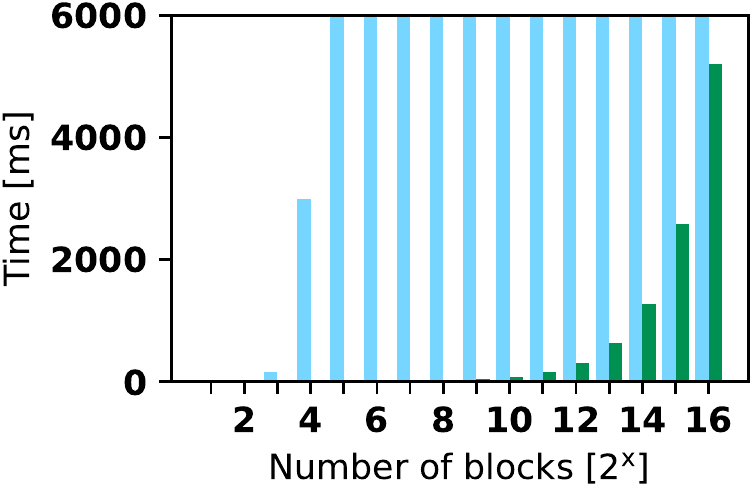} & \includegraphics[width=40mm]{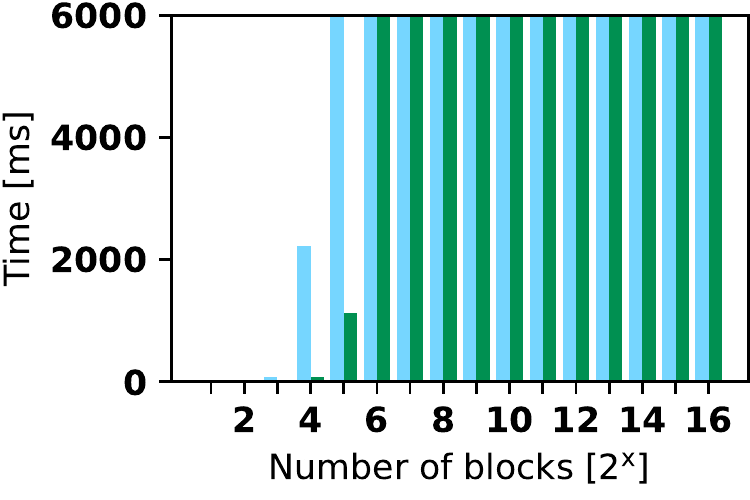} \\
\end{tabular}
\vspace*{-0.2cm}
\caption{Comparison of Naive Strategy and Optimized Strategy for Overlap Detection and Elimination.}
\vspace*{-0.6cm}
\label{fig:overlap}
\end{figure*}

Still, we are able to compare the workflow of both systems. To do so, we replicated the pipeline of ClimateSpark within \system{}. Precisely, we first hard-coded the list of files as well as the subarrays and variables to load for each file to simulate the effects of ClimateSpark's spatiotemporal index. Note that we do not measure the overhead that building a spatiotemporal index would normally imply. 
The resulting RDD maps tuples consisting of variable name, timestamp and spatial bounding box to entire subarrays, representing the values.
This RDD is then transformed into a row-wise representation following the example of \smalltt{queryPointTimeSeries} from the ClimateSpark repository~\cite{lit:climatespark-github}. The row-wise representation is converted to a SparkSQL \smalltt{DataFrame}, which includes five columns: The name of a variable, the three-dimensional coordinates, and the associated value.
If the query requires multiple variables, each value is stored in a separate row.

Again, we evaluate both pipelines at two applications from atmospheric physics.
In Figure~\ref{fig:climatespark:wind}, we query the maximum wind speed within a contiguous 24-hour window on ERA5\_Sur.
We vary the extent of the window along the latitude dimension.
Within the dataset, wind speeds are represented by two perpendicular components, so the total wind speed has to be computed from separate variables through a UDF.
However, since the ClimateSpark pipeline does not produce a DataFrame in which both components are in the same row, we need to perform a self-join first.
The execution time overhead associated with this self-join is apparent in the results.
Note that this self-join operation is not necessary in \system{} because all values belonging to the same coordinate are always grouped into a single row.

In Figure~\ref{fig:climatespark:oceans}, we compute the maximum convective precipitation over the North Atlantic and North Pacific regions on ERA5\_Pre.
We vary the number of files in our base dataset up to an entire month of data.
Since ClimateSpark only supports a single convex block to be loaded, we load the smallest possible block that fully contains our regions of interest, then filter the rows afterwards.
The difference in execution time shown in Figure~\ref{fig:climatespark} is likely due to this issue.

\vspace*{-0.2cm}
\subsection{Experiment: Overlap Detection and Elimination Strategies}
\vspace*{-0.1cm}
Lastly, we evaluate the two proposed strategies for overlap elimination, which is required to handle non-convex predicates.
Note that since the size of the blocks does not impact the execution time of both strategies, we only vary the number of blocks~$n$ from $2$ to $2^{16}$ and the number of dimensions~$d$ from $1$ to $4$ in each experiment.
We provide four workloads, where the first two workloads are rather common in practice. The last two workloads are artificially designed to maximize the number of distinct interval boundaries.
\textbf{WL\_aligned}:~$n$~blocks sharing a common origin and extent along every dimension except dimension 0, where blocks are arranged next to each other. Two adjacent blocks overlap with a probability of 50\%.
\textbf{WL\_misaligned}:~Similar to WL\_aligned, but every block has a 50\% chance to be shifted by half its extent along any dimension except 0.
\textbf{WL\_diagonal}:~\mbox{$n-1$}~blocks arranged diagonally inside a single large block.
\textbf{WL\_centered}:~$n$~overlapping blocks sharing a common center where block $i$ has greater extent along dimension 0 but smaller extent along all other dimensions when compared to block $i-1$.

Figure~\ref{fig:overlap} shows the average execution time over five runs.
As we can see, both strategies scale linearly with the number of blocks for WL\_aligned. This behavior is expected since each additional block introduces exactly two new interval boundaries.
The naive strategy is slightly slower because every block has to be decomposed explicitly before being merged, whereas the optimized strategy has to perform merges only.
WL\_misaligned introduces additional interval boundaries due to some blocks being misaligned, thus further increasing the gap between the two strategies.
For WL\_diagonal, the naive strategy splits the large block into $(2n-1)^d$ smaller blocks. This exponential growth with increasing dimensionality drastically hurts its performance. The optimized strategy slices and reassembles the large block one interval at a time, so the intermediate results are much smaller.
For WL\_centered, the naive strategy generates lots of copies of sub-blocks close to the center, up to $n$ copies for the region where all blocks overlap. In contrast, the optimized strategy eliminates these redundant blocks immediately in the interval merge step.

Overall, we can see that while the naive strategy quickly becomes infeasible with increasing $n$, the optimized strategy remains usable for small high-overlap scenarios. Thus, by default, \system{} uses the optimized strategy.

%
%

\vspace*{-0.3cm}
\section{Conclusion}
\vspace*{-0.2cm}
In this work we proposed~\system{}, a system which allows researchers from Earth sciences to formulate their analytical tasks in convenient SQL \textit{while} being able to rely on several automatic optimizations in the background. These optimizations are tailored towards observational datasets in NetCDF format and enable a maximal amount of pruning of data directly at the data source -- even for queries, that formulate complex non-convex predicates. We show that the performance of~\system{} scales gracefully with the horizontal and vertical selectivity of the queries and that \system{} outperforms the comparable state-of-the-art pipeline by up to a factor of 6x. 

\textbf{Acknowledgements}:
This work contributes to the project  “Big Data in Atmospheric Physics (BINARY)”, funded by the Carl Zeiss Foundation (grant P2018-02-003).

\bibliographystyle{IEEEtran}
\bibliography{bib-refs}

\end{document}